\documentclass[a4paper,onecolumn,10pt]{article}
\usepackage[utf8x]{inputenc}

\usepackage{graphicx}
\usepackage{upgreek}
\usepackage{dcolumn}
\usepackage{latexsym}
\usepackage{color}
\usepackage{hyperref}
\usepackage{amsmath}
\usepackage{anysize}
\usepackage{fancyhdr}
\usepackage{fancybox}
\usepackage{titlesec}
\usepackage{hyperref}
\usepackage{natbib}
\usepackage{enumitem}

\title{Voxel datacubes for 3D visualization in Blender}
\author{\rm Matías Gárate$^{1,2}$\\
}
\date{\small
    $^1$Instituto de Astrofísica, Pontificia Universidad Católica de Chile, 7820436 Santiago, Chile\\%
    $^2$Millennium Nucleus \textquotedblleft Protoplanetary Disks in ALMA Early Science\textquotedblright, Santiago, Chile \\[1ex]%
    Email: migarate@uc.cl\\
    Accepted in the Publications of the Astronomical Society of the Pacific
}

\begin{document}
\maketitle

\section*{Abstract} 

The growth of computational astrophysics and complexity of multidimensional datasets evidences the need for new versatile visualization tools for both analysis and presentation of the data. In this work we show how to use the open source software Blender as a 3D visualization tool to study and visualize numerical simulation results, focusing on astrophysical hydrodynamic experiments.
With a datacube as input, the software can generate a volume rendering of the 3D data, show the evolution of a simulation in time, and do a fly-around camera animation to highlight the points of interest.
We explain the process to import simulation outputs into Blender using the Voxel Data format, and how to set up a visualization scene in the software interface.
This method allows scientists to perform a complementary visual analysis of their data, and display their results in an appealing way, both for outreach and science presentations.\\

\noindent \textbf{Keywords}: methods: miscellaneous -- methods: numerical\newline

\section{Introduction}
The amount of data available in astronomy grows continuously in size and complexity.
This creates the need to develop more effective visualizations techniques to analyze and present the data.\\ \indent 
Typically, astronomers work analyzing datacubes, which are simply multidimensional arrays of values, such as the physical fields of hydrodynamic simulations or integral field spectroscopy data. It is easy to display the complete information of 1D arrays or 2D matrices using lineplots and colormaps. However, a problem appears for 3D datacubes, when only a single projection or a sequence of layers are displayed to visualize the dataset. This procedure comes with a reduction of the dimensionality of the data, and can cause to miss part of the information in the visual analysis.\\ \indent 
3D visualization and Volume Rendering are now alternatives for astronomers to display 3-dimensional data \citep{barnes08}. Current softwares and libraries allow the user to do it in real time with different levels of interactivity. 
Viewers like FRELLED\footnote{\href{http://www.rhysy.net/frelled-archive.html}{www.rhysy.net/frelled-archive.html}} \citep{taylor15} allow to load FITS datacubes and manipulate them interactively through rotations, scaling, and mask selection. Alternatives like Virtual Reality immersions go one step forward, and let the user get inside the data to observe its features \citep{ferrand16}.
Using these novel techniques not only allow to produce an attractive outreach material, but also provide an additional perspective of the complete dataset while doing scientific analysis along with the traditional methods\citep{goodman12}.\\ \indent 
In this work we provide a procedure to generate a Volume Rendering from astronomical data (from numerical simulations, in particular) in the 3D graphics software Blender\footnote{\href{https://www.blender.org/}{www.blender.org}}, that can be used to perform a qualitative visual analysis, and to create presentation material.
Blender is free and open source, and in the recent years has shown potential as an astronomy visualization tool. \cite{kent13} presents a series of tutorials\footnote{\href{http://www.cv.nrao.edu/~bkent/blender/}{www.cv.nrao.edu/$\sim$bkent/blender/}} showing the software features and possible applications like: loading as halo points a galaxy catalog or a particle simulation, use the mesh modeling tools to recreate an asteroid, and also a demonstration of volume rendering using an image sequence. The library AstroBlend\footnote{\href{http://www.astroblend.com/}{www.astroblend.com/}}, developed by \cite{naiman16}, provides further tools to import and display different types of astronomical data in the interactive 3D environment of Blender, such as 3D contours and also particle simulations using halos with a colorscale.\\ \indent 
A datacube can be imported into Blender through the Voxel Data format, which is the file used to generate volumetric renderings.
By importing a simulation, for example, the software can display in a 3D space one of the physical properties of interest (density, temperature, etc) using a colorscale, and show its evolution in time. Using the Blender functions it is possible adjust the colorscale interactively to highlight different features, incorporate a fly-around camera animation to zoom in through the relevant regions of the datacube, and finally produce an animation to use as outreach material.\\ \indent 
The use of Volume Rendering provides a smooth and continuous display of the whole dataset, in contrast with visualizations using particles or contours. Since Blender colorscale also provides control over the transparency, it is also possible to explore the different depths of the volume without reloading additional data.\\ \indent
This article is meant to serve as a guideline for the user to convert its own data into the Voxel Data format, learn the steps to set up a Blender scene, and render an image or a movie.
Section \ref{sec_voxel} explains the format of the voxel data files that can be used as an input to Blender. 
Section \ref{sec_convert} provides simple algorithms to convert some common simulation data formats into voxel data. 
In section \ref{sec_setup} it is described the setup process in Blender from the initial scene, through loading the Voxel Data, and finally the rendering process. 
Section \ref{sec_examples} displays examples using Gadget2\citep{springel05} and FARGO3D\citep{benitez16} simulations rendered in Blender.\\ \indent 
To complement this article, the reader can watch the tutorial series: 
``Blender \& Astronomy Tutorial. Using Voxel Data for 3D visualization''\footnote{\href{https://www.youtube.com/watch?v=zmY\_mn6Ue2g\&list=PLjFmkbKBKd0WuhhbR2MGPXZyWW7TPcNSw}{www.youtube.com/watch?v=zmY\_mn6Ue2g\&list=PLjFmkbKBKd0WuhhbR2MGPXZyWW7TPcNSw}}, 
which covers the whole process described above with real time demonstrations in Blender. 
Through the article these videos will be referred as TUT:\#. 
All the material used for this work is available in the github repository: 
``Blender Bvoxer''\footnote{ \href{https://www.github.com/matgarate/Blender_Bvoxer}{github.com/matgarate/Blender\_Bvoxer}},
which includes the Blender scenes, scripts to generate voxel data files, and simulation data provided by J. Cuadra and S. Perez to test this visualization techniche. Through the article these codes will be referred as GIT:FolderName.

\section{Understanding Voxel Data in Blender} \label{sec_voxel}
This section will explain the key concepts related to the Voxel Data, which is the link between the astrophysical data, and the 3D visualization in the Blender scene. 
The Voxel Data is a binary datacube file that will serve as input for the visualization. 
Although the Voxel Data can be generated from a sequence of images (as we will discuss later in this section), we prefer (and recommend) to use a single binary file to store the information. 
This approach allows to represent the datacube as a single array (which is convenient when performing operations over the data), it easier and faster to write a single binary file than multiple images, and finally, it allows to store all the information in a compact format, rather than in a directory.
To complement this section the reader can refer to the tutorial series TUT:1 and github repository GIT:Example, where there are demonstrations for generating a simple Voxel Data file from a function $f(x,y,z)$.\\ \indent 
To visualize 3-dimensional data it is necessary to define a finite volume within the Blender scene, this is done by creating a Cube object
that will be the domain for the visualization. Once the Voxel Data file is assigned to this domain, the 3D information stored in the file will
be rendered in the Blender scene within the boundaries of the Cube. 
Further technical details of the setup, such as the material and texture settings, will be addressed later in section \ref{sec_setup}.\\ \indent 
The voxel file represents a datacube subdivided in the three cartesian axes $(x, y, z)$. For each discrete cartesian coordinate $(i, j, k)$ inside the volume there is one voxel value $V(i, j, k)$. Additionally, it is possible to include multiple snapshots (frames) in the same file. This allows to show the evolution of the datacube in time.
Therefore, the Voxel Data is defined by:
\begin{itemize}
 \item The resolution numbers $(n_x, n_y, n_z)$, that indicate the number of subdivisions of the domain in each axis.
 \item The frame number $n_f$, that indicates the total number of frames contained in the file.
 \item $(n_x\times n_y\times n_z)\times n_f$ values, which describe the astrophysical data that will be visualized inside the domain.
\end{itemize}
By the time of writing this article, Blender can receive the Voxel Data as an input in three different file formats: 
Blender Voxel, 8bit Raw and Image Sequence. 
The following subsections will describe each format, and point their advantages(+) and disadvantages(--).\\ \indent 
The guidelines and notation provided in this article to write both Blender Voxel and 8bit Raw file formats are based on the description and examples available at Pythology Blogspot\footnote{\href{http://pythology.blogspot.cl/2014/08/you-can-do-cool-stuff-with-manual.html}{pythology.blogspot.cl/2014/08/you-can-do-cool-stuff-with-manual.html}}.
Also, \cite{kent13} (section 4.1) provides a brief description about how to use the Image Sequence format and a complete example on his website\footnote{\href{http://www.cv.nrao.edu/~bkent/blender/tutorials.html}{www.cv.nrao.edu/$\sim$bkent/blender/tutorials.html}}.
\subsection{Blender Voxel format} \label{subsec_bvox}
The Blender Voxel format consists in a binary file that has $4\times 32\rm bit$ integers in the header, corresponding to $(n_x, n_y, n_z,n_f)$, followed by the $(n_x\times n_y\times n_z\times n_f) \times 32\rm bit$ floating point values that represent the actual data. For this format the values must be normalized to a [0,1] interval.
The order to write the full datacube after the header is:
\begin{enumerate}
 \item Frame by frame, where each frame has $n_z$ layers.
 \item Layer by layer, where each layer has $n_y$ lines.
 \item Line by line, where each line has $n_x$ values.
 \item Value by value, until completing the $n_x$ values.
\end{enumerate}
The following example pseudo-code illustrates the order in which the file is written:
\begin{verbatim}
write([nx,ny,nz,nf])
for t from 0 to nf:		
 for k from 0 to nz:		 
  for j from 0 to ny:		
   for i from 0 to nx:		 
    write(Data[i,j,k,t])
\end{verbatim}
\begin{itemize}
 \item (+) Allows high dynamic range for the data values by using 32bit floats.
 \item (+) Includes a header. It only requires to input the file into Blender.
 \item (+) Allows multiple frames.
 \item (--) High memory usage because of using 32bit float.
\end{itemize}
We want to remark that the high memory usage may be specially troublesome when multiple frames are stored in the Voxel Data file, since it is not possible create arbitrarily large files. 
For large number of frames it may be more convenient to store each frame in a separate Voxel file, and then use the python API of Blender to switch between frames. This will be discussed again in section \ref{sec_setup_tips}.

\subsection{8bit Raw format} \label{subsec_8bit}
This format is a binary file of $(n_x\times n_y\times n_z\times n_f) \times 8\rm bit$ integer values. The values for this format should normalized to a [0,255] interval. The order to write the file is the same as the used for the Blender Voxel format, the only difference is the absence
of the header and the size of the values. For this format the $(n_x, n_y, n_z)$ values must be written as an input in the Blender interface. 
The value of $n_f$ is calculated internally by Blender based on the file size and the previous three values.
\begin{itemize}
 \item (--) Low dynamic range for the data values.
 \item (--) Does not include a header. It is necessary to input the resolution manually in Blender.
 \item (+) Allows multiple frames.
 \item (+) Memory efficient because of using 8bit integers.
\end{itemize}

\subsection{Image Sequence}
The Image Sequence format allows to construct a datacube using multiple images as layers. The format consists in a directory that contains $n_z$ 
image files (png, for example) of $n_x\times n_y$ pixels. 
If the user already has the tools to save his data as a set of images, this would be the most straightforward method to construct the Voxel Data.\\ \indent 
The total number of images must be specified in the interface along with the image \#1. The ($n_x$, $n_y$) values are obtained from the files. 
Those files must be numerated to guarantee that the datacube will be build in the correct order. 
For example, if there are 100 images a valid numeration would be: name001.png,..., name100.png.
\begin{itemize}
 \item (+) Convenient if the images can be easily generated in order to build the datacube. This would save extra coding.
 \item (+) Does not need a header. Blender computes the resolution automatically from the number of images and their dimensions.
 \item (--) Only support a single frame.
 \item (--) Requires multiple files.
\end{itemize}
The memory usage of this method will depend on the image number of channels and their dynamic range. 
However, notice that Blender will interpret all images as monochromatic, so using anything but a grayscale is unnecessary.
The dynamic range of the grayscale will determine the level of detail. For example, an 8bit image will provide greater detail than a 1bit image at the cost of memory usage. We have tested this method with 1bit, 8bit grayscale images, and also with 24bit (RGB) and 32bit (RGBA) images. Further information can be found in \cite{kent13}.

\section{Create Voxel Data from simulation outputs}\label{sec_convert}
As mentioned in the previous section, the Blender Voxel and 8bit Raw inputs take the form of datacubes.
By the time of writing this article, Blender can only subdivide the visualization domain in the cartesian axes (x, y, z). 
Therefore, the algorithm to construct the voxel file needs to fill each grid cell of this datacube from the simulation data.\\ \indent 
For a simulation that uses an evenly spaced cartesian grid, writing the voxel data is quite straightforward, as the only thing required to do is to match $(n_x, n_y, n_z)$ with the resolution of the simulation, and scale the data values to the appropriate range before writing the file in the Blender Voxel or 8bit Raw format. 
However, if the simulation is done using a cylindrical or spherical grid, a logarithmic scale, or if it is a particle simulation (like N-Body or SPH), it is necessary to transform the space of the original data into a cartesian datacube.\\ \indent 
In this section we propose methods to generate a cartesian datacube from a grid simulation in spherical coordinates, and also from a particle simulation.
As it is impossible to address all the possible data formats and methods we expect that these two should serve as good examples that the user can adjust to his/her own needs.\\ \indent 
The tutorial video TUT:3 (and the second half of TUT:4) also explain these methods by going through the example scripts, however these videos are aimed to people with little or none programming skills. An experienced programmer will find more convenient to read the following subsections and look the referenced codes straight away.

\subsection{Spherical grid-based simulation}\label{sec_convert_grid}
Converting an spherical grid into a cartesian datacube requires to fill each voxel $(i, j, k)$ (where $i\in[0, n_x-1]$, $j\in[0, n_y-1]$, $k\in[0, n_z-1]$) 
using the spherical coordinates $(r,\phi,\theta)$, which are the radial, azimuthal and co-latitude directions in the simulation, and the physical
properties $F$ of each grid cell, such as density, energy, temperature, etc.
Notice that the spherical coordinates should indicate the middle of the grid cell, and not the borders.\\ \indent 
We propose the following algorithm to perform the transformation:

\begin{enumerate}[label*=\arabic*.]
 \item First transform the $(r,\phi,\theta)$ coordinates into their cartesian equivalent $(x, y, z)$.
 \item Define the cartesian boundaries of the datacube using the minimum and maximum $(x, y, z)$ values from the previous step.
 \item Define a resolution $(n_x, n_y, n_z)$ for the voxel datacube.
 \item For each voxel position $(i, j, k)$ obtain its cartesian equivalent ($x_i, y_j, z_k$), using
 the following transformation $i\rightarrow x_i= x_{min} + i\cdot(x_{max}-x_{min})/(n_x-1)$. 
 Repeat this for $j\rightarrow y_j$ and $k\rightarrow z_k$.
 \item Check if the values $(x_i, y_j, z_k)$ are within the spherical boundaries of the simulation. If they do, proceed to the next step. If they do not, assign the voxel $V(i, j, k)=0.0$. This is expected to happen since a spherical grid cannot fill completely the borders of a cartesian grid, or the regions inside the inner radial boundary of the simulation.
 \item The value of each voxel $V(i, j, k)$ in the datacube with position $(x_i, y_j, z_k)$
 will be defined by the $N$ closest grid cells of the simulation and their physical properties $F(C_a)$, with $C_a$ the $a^{th}$ closest cell neighbor.
 We present two alternatives for determining $V(i, j, k)$. 
 \begin{enumerate}[label*=\arabic*.]
  \item For $N=1$ use the closest neighbor as the only contribution to the voxel value. Then, $V(i, j, k)= F(C_1)$. 
  \item For $N=8$ it is possible to use a trilinear interpolation if the 8 positions $(x, y, z)$ of the neighbors
  form a box that contains the $(x_i, y_j, z_k)$ coordinate of the voxel. Then, 
  \begin{equation}
  V(i,j,k)=\sum\limits_{a=0}^{N=8} w(C_a)\cdot F(C_a), 
  \end{equation}
  where $w(C_a)$ is the weight for the trilinear interpolation of the grid cell $C_a$.
  (This is the method used in the example shown in Figure \ref{ExampleA}).
 \end{enumerate}
 \item After having all the voxel values, rescale them to be in the interval $[0, 1]$ or $[0, 255]$, depending if the voxel format is Blender Voxel
 or 8bit Raw, respectively.
 Notice that in the step 5 a value of 0.0 was assigned to the voxels outside the simulation boundaries, so it may be necessary to force them again back to 0.0 
 for the visualization if the property $F$ allowed negative values.
\end{enumerate}
An implementation of this algorithm is available in the github repository GIT:Fargo3DVoxelizer, where FARGO3D outputs are converted from an 
spherical grid to Blender Voxel format. Both alternatives for steps 6.1 and 6.2 are available in the repository. While the former is enough for the visualization, it may cause discontinuities in the sampling. On the other hand, the latter solves this problem and samples the cartesian datacube smoothly.

\subsection{Particle-based simulation}\label{sec_convert_particle}
The process to generate a datacube from a particle simulation can be divided in two steps: First, to define a cartesian grid and find the position of the particles on it, and second, to use the particles in each grid cell to compute its voxel value $V(i, j, k)$.\\ \indent
For a simulation with $N$ particles, we have their position, velocities and physical properties $F$. The latter can be masses and sizes for N-Body simulations, or density, thermal energy and other fluid properties for hydrodynamic simulations like SPH \citep{monaghan92}.\\ \indent 
We propose the following algorithm to convert an array of particles into an evenly spaced datacube:
\begin{enumerate}[label*=\arabic*.]
 \item Define the boundaries of the simulation using the minimum and maximum values of the particles positions in each axis $(x, y, z)$. 
 \item Define a resolution $(n_x, n_y, n_z)$ that is convenient for the particle count and available memory. 
 \item Transform the $x$ positions of the particles from the range [$x_{min}$, $x_{max}$] to the range [$0, nx$] using the linear transformation
 $x \rightarrow x_v=\frac{x-x_{min}}{x_{max}-x_{min}}n_x$. Repeat for the y and z coordinates.
 \item Identify if a particle $P$ is inside (or should influence) a certain grid cell $(i, j, k)$, and its contribution to the voxel value  $V(i, j, k)$. For this step, we propose the following alternatives:
 \begin{enumerate}[label*=\arabic*.]
  \item Simple average estimation.
  Truncate the transformed positions to their integer forms. A particle $P$ is inside a certain grid cell if the coordinates   $(int(x_v), int(y_v), int(z_v))$ match the $(i, j, k)$ indexes of the cell in the datacube. The value of each voxel will be given by 
  \begin{equation}
    V(i, j, k) = \sum\limits_{a=0}^{N_{ijk}} \frac{1}{N_{ijk}}\cdot F(P_a),
  \end{equation}
  where $P_a$ are the $N_{ijk}$ particles inside a grid cell $(i,j,k)$, 
  and $F(P_a)$ is the physical property chosen for visualization of the $P_a$ particle.
  (This is the method used in the example shown in Figure \ref{ExampleB}).
  \item SPH interpolation.
  Using the interpolation method implemented in SPLASH\footnote{\href{http://users.monash.edu.au/~dprice/splash/}{users.monash.edu.au/$\sim$dprice/splash/}} (\cite{price07}, section 4.1 and 4.2), the value of each voxel is given by:
  \begin{equation}
    V(i, j, k) = \sum\limits_{a=0}^{N} \frac{m_a}{\rho_a} \cdot F(P_a) W(r, h), 
  \end{equation}
  where $W$ is the SPH kernel function, r is the distance between the particle $P_a$ and the center of the grid cell $(i, j, k)$, $m_a$ and $\rho_a$ are
  the mass and density of the particle, and $h=\max(h_a,\Delta/2)$ is the maximum between the smoothing length of the particle 
  and the half width of the grid cell.
  \item Other possible methods could include using the N-nearest particles to influence a grid cell, and use a customized weight functions
  instead the SPH kernel, such as power laws of the distance. The value of the voxel can be computed in an analogous way as in the previous methods.
  \end{enumerate}
  These procedures can be optimized by iterating over the particles, and adding their contribution to the voxels that are within their range of influence, rather iterating over all the voxels and looking for the contribution of all the N particles \citep{price07}. 
 \item The last step is to rescale the whole voxel datacube to have values in the interval $[0, 1]$ or $[0, 255]$, depending if the voxel format 
 is Blender Voxel or 8bit Raw, respectively. 
\end{enumerate}
An implementation of this algorithm is available in the github repository GIT:SPH. The example code reads a particle list, and generates a Blender Voxel file using the density field of the particles. The approach of step 4.1 is used for simplicity, but the interpolation recipe described by \cite{price07} should be implemented for a more accurate result, specially if the size of the grid cells is smaller than the relevant smoothing lengths of the simulation.

\section{Blender set-up for visualization}\label{sec_setup}
\begin{figure}[t!]
\begin{center}
\includegraphics[width=120mm]{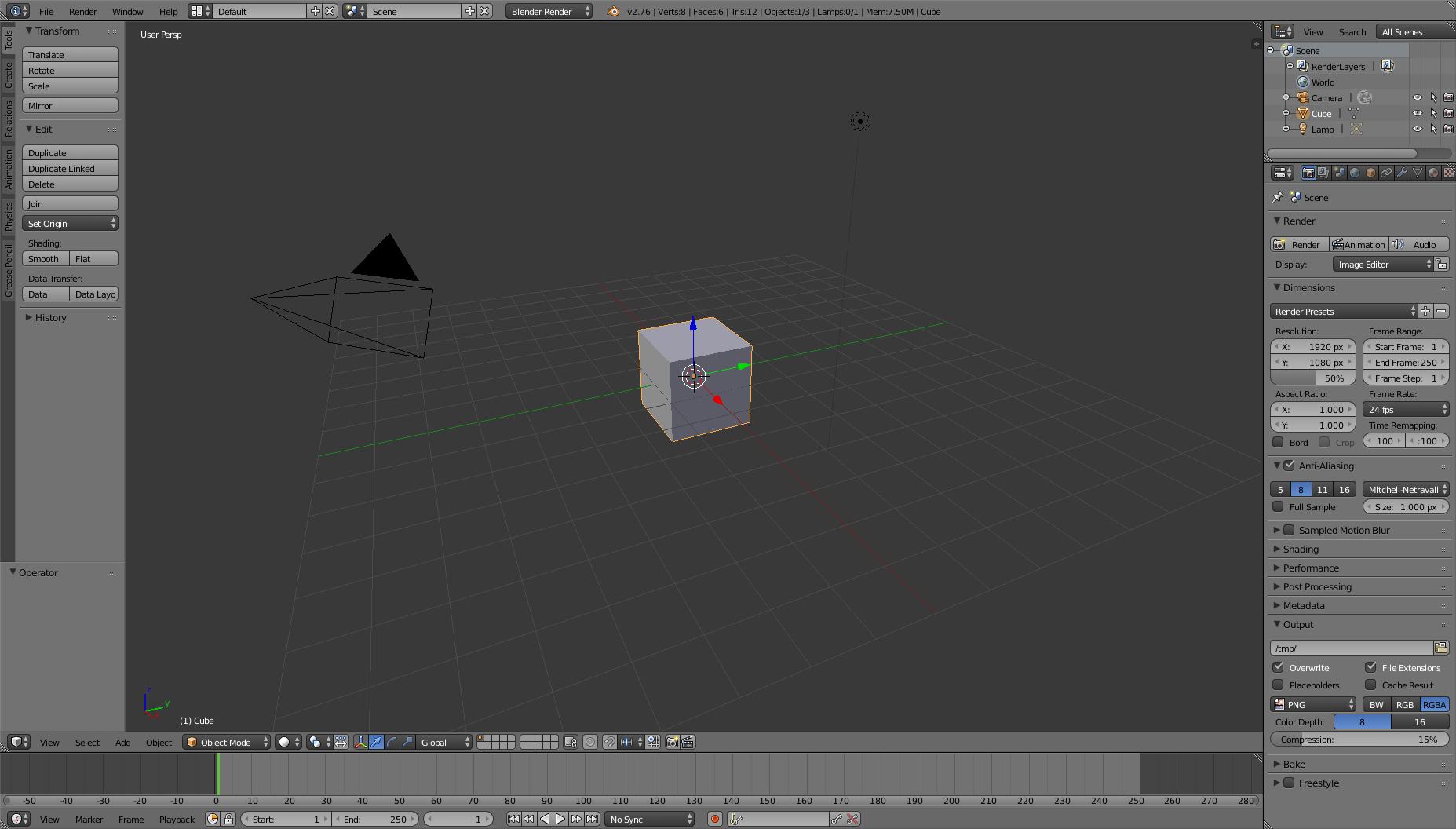}
 \caption{This is the default Blender scene, which includes a default \textbf{Cube}, a \textbf{Camera} and a \textbf{Lamp}. With the \textbf{Cube} selected, 
 it is possible to access to the \textbf{Material} and \textbf{Texture} properties in the right panel of the interface.}
 \label{Set1}
\end{center}
\end{figure}

In order to produce a Volume Rendering, the light needs to pass through the defined volume, be absorbed, scattered, and transmitted by interacting with the voxels according to their properties (density, emission, etc), until it reaches the camera.
The concept is similar to the radiative transfer process in astrophysics, although interactions are computed in the color space, rather than in wavelenght.\\ \indent
This section will cover the overall steps required to load a Voxel Data file into Blender, set up the material, render an image, and also create a fly-around camera animation. This procedure is also demonstrated in our tutorial series TUT:2 in real time.
\cite{kent13} also covered the setup of a Blender scene, going through the steps of a standard workflow.\\ \indent 
Before starting, it is advised to take a look at the basic Blender commands, just to be able to navigate quickly through the interface. Some pages that cover the basics are: Blender 3D Design Course - Lesson 1\footnote{\href{http://gryllus.net/Blender/Lessons/Lesson01.html}{gryllus.net/Blender/Lessons/Lesson01.html}} by Neal Hirsig, and also the Blender Basics course\footnote{\href{https://cgcookie.com/course/blender-basics/}{cgcookie.com/course/blender-basics/}} from CG Cookie. To learn about any specific feature of the software the reader can refer to the online Manual\footnote{\href{https://www.blender.org/manual/}{www.blender.org/manual/}}.\\ \indent 
After opening Blender, the default scene will be displayed as shown in Figure \ref{Set1}.

\subsection{Setup the 3D texture}\label{sec_setup_3dtex}
\begin{figure}[t!]
\begin{center}
\includegraphics[width=140mm, height=130mm]{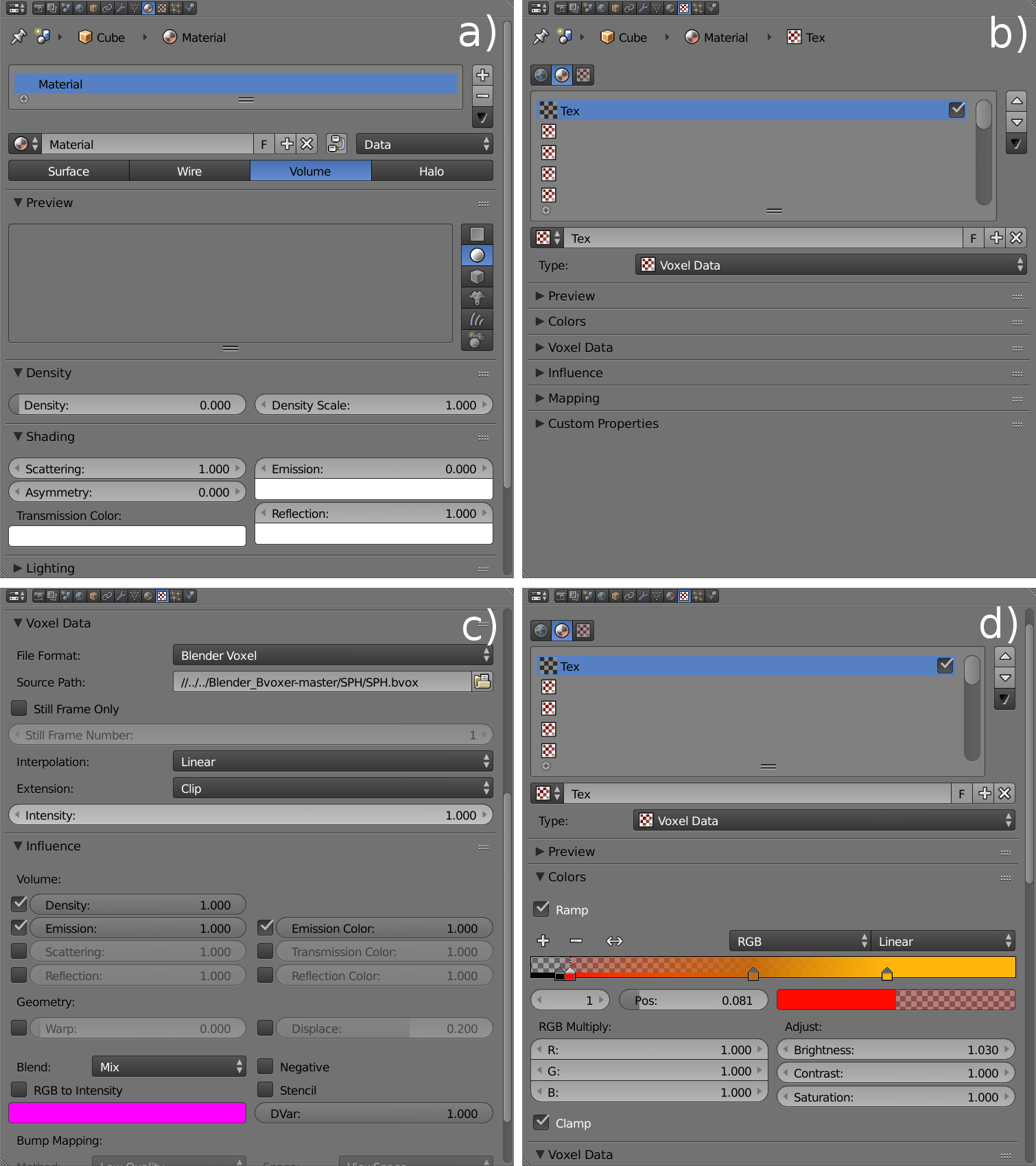}
 \caption{Steps to import the Voxel Data and set up the 3D texture. 
 (a) Set the \textbf{Material} to \textbf{Volume} and \textbf{Density} to \textbf{0.0}. 
 (b) Set the \textbf{Texture Type} to \textbf{Voxel Data}.
 (c) In the \textbf{Voxel Data} panel, set file format to \textbf{Blender Voxel} or \textbf{8bit Raw} (this last one requires to specify the dimensions of the datacube), and load the datacube file. In the \textbf{Influence panel}, set the \textbf{Density}, \textbf{Emission}, and \textbf{Emission Color} to \textbf{1.0}. 
 (d) In the \textbf{Color} panel, activate the \textbf{Color Ramp}, then adjust the sliders to customize your colorscale, adjusting the \textbf{Brightness} and \textbf{Contrast} may help to highlight the features of the datacube.}
 \label{Set2}
\end{center}
\end{figure}

Here are listed the steps to set the 3D material, starting from the default scene. The default \textbf{Cube} object will act as domain for the Voxel Data as mentioned in section \ref{sec_voxel}. For the complete procedure we will use the default internal engine \textbf{Blender Render}, which allows to use customized voxel files as input.
\begin{enumerate}
 \item Right-clicking on the default \textbf{Cube} gives access to the object properties, including the \textbf{Material} and \textbf{Texture} 
 properties. Select the \textbf{Material} property, set the type to \textbf{Volume} and \textbf{Density = 0.0} (see Figure \ref{Set2} (a)). The integration of the light rays through the volume can be customized in the \textbf{Integration} panel through the \textbf{Step Size} and the \textbf{Depth Cutoff}, both parameters can be adjusted to improve the accuracy at the cost of rendering time. More details about the other Volume Rendering settings can be found in its respective section in the manual\footnote{\href{https://www.blender.org/manual/render/blender_render/materials/special_effects/volume.html}{www.blender.org/manual/render/blender\_render/materials/special\_effects/volume.html}}.
 \item Select the \textbf{Texture} property and set the type to \textbf{Voxel Data} (see Figure \ref{Set2} (b)). In the \textbf{Voxel Data} panel select the file format to one of the discussed in the section \ref{sec_voxel} (see Figure \ref{Set2} (c)). 
 If the datacube used contains only a single frame ($n_f = 1$), then check the \textbf{Still Frame Only}, and set the \textbf{Still Frame Number = 1} to display the frame.
 For $n_f > 1$ leave \textbf{Still Frame Only} unchecked. In this case the \textbf{Current Frame} value of the Blender \textbf{Timeline} will determine  which frame of the datacube should be displayed.
 \item In the \textbf{Influence} panel check the \textbf{Density}, \textbf{Emission} and \textbf{Emission Color} fields, and set them all to \textbf{1.0} (see Figure \ref{Set2} (c)).
The \textbf{Influence} panel determines which parameters of the volume will be defined from the \textbf{Voxel Data} input. The \textbf{Density} field defines the opacity of the volume, the \textbf{Emission} defines its brightness, and \textbf{Emission Color} allows to use a colorscale that can be customized in the \textbf{Color} panel, instead of a grayscale.

\item In the \textbf{Color} panel check the \textbf{Ramp} option (see Figure \ref{Set2} (d)). The ramp is a function that receives the values of the \textbf{Voxel Data} file as a position between \textbf{0.0} and \textbf{1.0}, and returns the color specified at that position (notice that it is also possible to control the transparency with the $\alpha$ channel, allowing see through or completely hide a range of values). Selecting the sliders allows to control their position in the ramp, and the color output. 
For the interpolation we suggest to start using the \textbf{RGB} color mode, and select the \textbf{Linear} method for the transition between consecutive sliders. The reader is encouraged to experiment with the other non-linear interpolation methods available in order to find the colorscale that best suits the data. Finally, adjusting the \textbf{Brightness} and \textbf{Contrast} helps to highlight the features of the volume and to expose its internal structure. This last part is mostly a process of trial and error, and may depend on the resolution of the \textbf{Voxel Data} file.  
\end{enumerate}

\subsection{World, Camera and Render}\label{sec_setup_scene}
\begin{figure}[t!]
\begin{center}
\includegraphics[width=140mm, height=130mm]{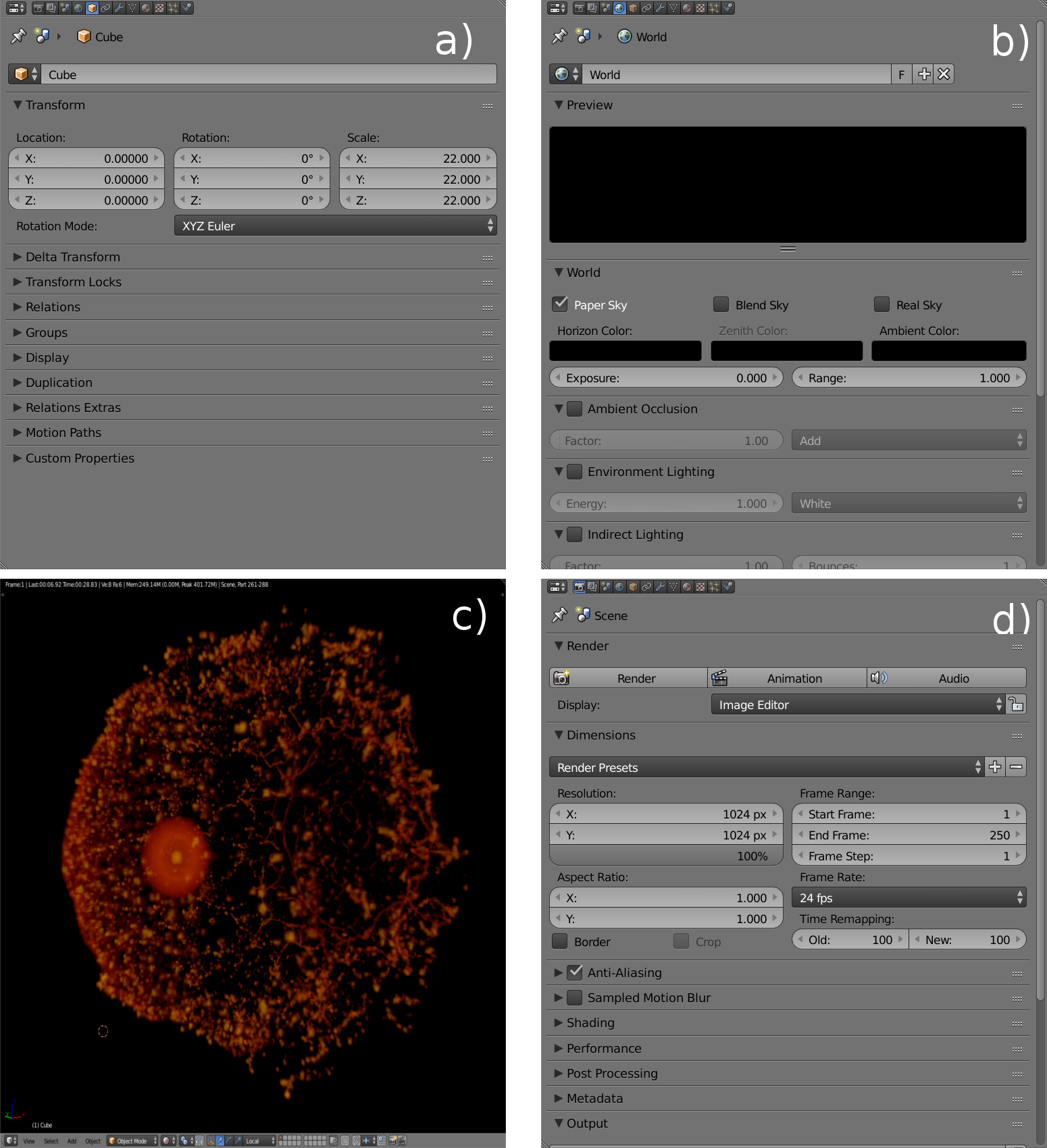}
 \caption{Steps to generate a quick render. (a) Rescale the \textbf{Cube} object in the \textbf{Object} properties. 
 (b) In  \textbf{World} properties, set the \textbf{Horizon} color to black.
 (c) In the 3D viewport, set the \textbf{Viewport Shading} to \textbf{Rendered} in order to previsualize the datacube.
 (d) In the \textbf{Render} properties select the image dimensions. Press \textbf{Render} to generate a single image, or \textbf{Animation} to 
 render a movie in the defined \textbf{Output} location.}
 \label{Set3}
\end{center}
\end{figure}

The following steps consist in the scene adjustments, positioning of the camera, correct the cube scaling, and other tweaks to improve the quality of the visualization.
\begin{enumerate} 
 \item Start by deleting the default \textbf{Lamp} object, since the datacube will illuminate by itself.
 \item It is recommended to scale up the \textbf{Cube} object in order to resolve better the volume (see Figure \ref{Set3} (a)). 
 It is advised to use a scale consistent with the geometry of the simulation, for example, the scale of a disk simulation in the z axis should be smaller than the scale of the x and y axes.  
 \item In the \textbf{World} properties it is convenient to set the \textbf{Horizon} color to black in order to increase the contrast between the Voxel Data and the background (see Figure \ref{Set3} (b)). 
 \item To have a preliminary visualization of the Voxel Data switch the \textbf{Viewport Shading} to \textbf{Rendered}. The viewport should look like Figure \ref{Set3} (c). This mode can be used to explore the data interactively, and to adjust the color, brightness and contrast in the \textbf{Texture} property in real time.
 \item To render a quick image, orient the \textbf{Camera} looking toward the Cube object, and position it at a distance such that the volume is contained inside the \textbf{Camera} field of view. There are many ways to do this and can be learned from every basic tutorial, but for now it is advised to use the command \textbf{Align Camera to View}. Then, select the resolution of the image in the \textbf{Render} properties and press \textbf{Render} (see Figure \ref{Set3} (d)).
 \item If the datacube contains multiple frames to generate an animation ($n_f > 1$), select the \textbf{Start Frame} and \textbf{End Frame} to go from 1 to $n_f$, remember to leave unchecked the \textbf{Still Frame Only} option in the \textbf{Texture} properties. Then, go to the \textbf{Output} panel in the \textbf{Render} properties, choose a destination and format for the movie file, and press the \textbf{Animation} button. 
\end{enumerate}

\subsection{Other setup tools}\label{sec_setup_tips}
Different setup tools can help to produce more appealing results. Here, we listed some of them, but since illustrating each step would greatly extend the length of this section, these will only be explained briefly. These are standard methods in Blender that can be found in many video tutorials, including tutorials TUT:2 and TUT:4.

\begin{itemize}
 \item Camera Fly-Around: The \textbf{Camera} can be constrained to move in time through a customized path, and kept pointing toward the object of interest. This allows to focus on the different regions of the simulation as it evolves in time, or simply go around all the details of a single frame of the datacube.\\ \indent 
 The overall procedure is: 
 \begin{itemize}
 \item Create a \textbf{Bezier Curve} or \textbf{Bezier Circle}, and go to \textbf{Edit Mode} in order to customize the animation path. 
 \item In the \textbf{Curve} properties check the \textbf{Path Animation} and set the number of frames needed to go through the whole path. 
 \item To animate the \textbf{Evaluation Time} set it to 0.0 at the starting frame, right click of the value slider and select \textbf{Insert Keyframe}. Repeat this last step at the final frame with a different value for the \textbf{Evaluation Time}. This will cause the \textbf{Evaluation Time} to change between the selected values during the animation. 
 \item Reset all the position and rotation values of the \textbf{Camera} object to 0.0. 
 \item Then, in the \textbf{Constraints} properties add the \textbf{Follow Path} constraint. Set the \textbf{Bezier Curve} as the \textbf{Target} object, the Y axis to point forward, the Z axis to point upwards. 
 \item Add the \textbf{Track To} constraint, and set the \textbf{Cube} as the target object, the -Z axis to point towards the \textbf{Cube}, and the Y axis to point upwards. 
 \item By pressing \textbf{Play} you should see the camera following the specified path.
 \item An optional method is to create an \textbf{Empty Object}, and use it as the focus of the \textbf{Camera} in the \textbf{Track To} constraint. Animating the position of this object allows to move the camera focus point through the animation.
 \end{itemize}
 
 \item Adding Halo-Points: As the data is related to astronomical subjects, it may be appealing to include a nice background if the audience is non-scientific. The \textbf{Halo Points} are ideal to simulate stars in the background of the scene.
	\begin{itemize}
	\item Add any \textbf{Mesh} object and switch from \textbf{Object Mode} to \textbf{Edit Mode}.
    \item Select all the vertices and use the \textbf{Merge at Center} command to collapse them into a single vertex. 
 	\item In the \textbf{Material} property, switch the type to \textbf{Halo}. Use \textbf{Size} and \textbf{Hardness} values to control the appearance of the halo.
 	\item To add multiple halo points at once, create an object like a \textbf{Sphere}, and in \textbf{Edit Mode} use the command \textbf{Delete Edges and Faces}. Repeat the material setup to produce as many halo points as vertices had the object.
	\end{itemize}

\item Compositor Post-Processing: The previous sections showed how to set up a \textbf{Color Ramp} for the datacube. 
 However, while the \textbf{Brightness} and \textbf{Contrast} values help to highlight the features of the data, these do not always produce a colorscale
 that favors the image as a whole. By switching from the \textbf{3D View} to the \textbf{Node Editor} it is possible to post-process the rendered image 
 for more appealing results. One useful node is the \textbf{Color Balance}, which corrects the image \textbf{Shadows}, \textbf{Midtones} and \textbf{Highlights}
 through the \textbf{Lift}, \textbf{Gamma} and \textbf{Gain} colors respectively.
 
 \item Switch Voxel Data files with python: It is also possible to automatize some of the tasks done in the interface using the \textbf{Blender API} for python scripting. As was noted in section \ref{sec_voxel}, storing a whole simulation in a single \textbf{Voxel Data} file can be inconvenient for long (or even moderate) simulations. In this case, it is easier to have one voxel file for each snapshot and perform the switch every frame through a script.\\ \indent 
 In the python script import the \textit{bpy} library (Blender internal library), and iterate over the frame count. Then, in each frame assign the corresponding voxel file to the texture, render the image, and save it. The individual images can be joined in a single video with the Blender \textbf{Movie Editor}, or with an external program. 
 An example script for this process is available in the github repository GIT:VoxelAnimation.
\end{itemize}

\section{Examples}\label{sec_examples}
\begin{figure}[t!]
\begin{center}
\includegraphics[width=85mm]{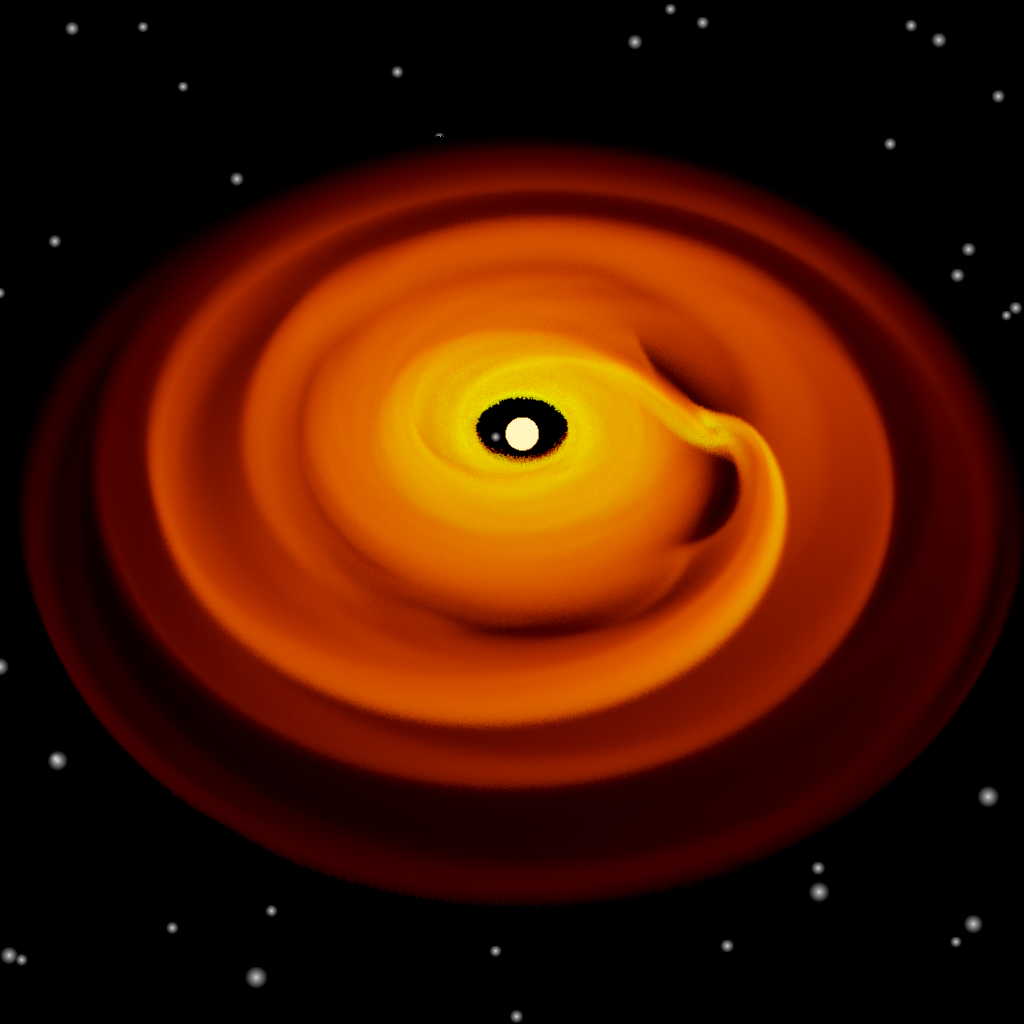}
 \caption{Protoplanetary disk with a massive planet carving a gap. Simulation data provided by S.Perez using FARGO3D, the output was converted from an spherical grid as described in the section \ref{sec_convert_grid}. The image was post-processed to brighten the colors, and the halo points were added to emulate surrounding stars.}
 \label{ExampleA}
\end{center}
\end{figure}

\begin{figure}[t!]
\begin{center}
\includegraphics[width=85mm]{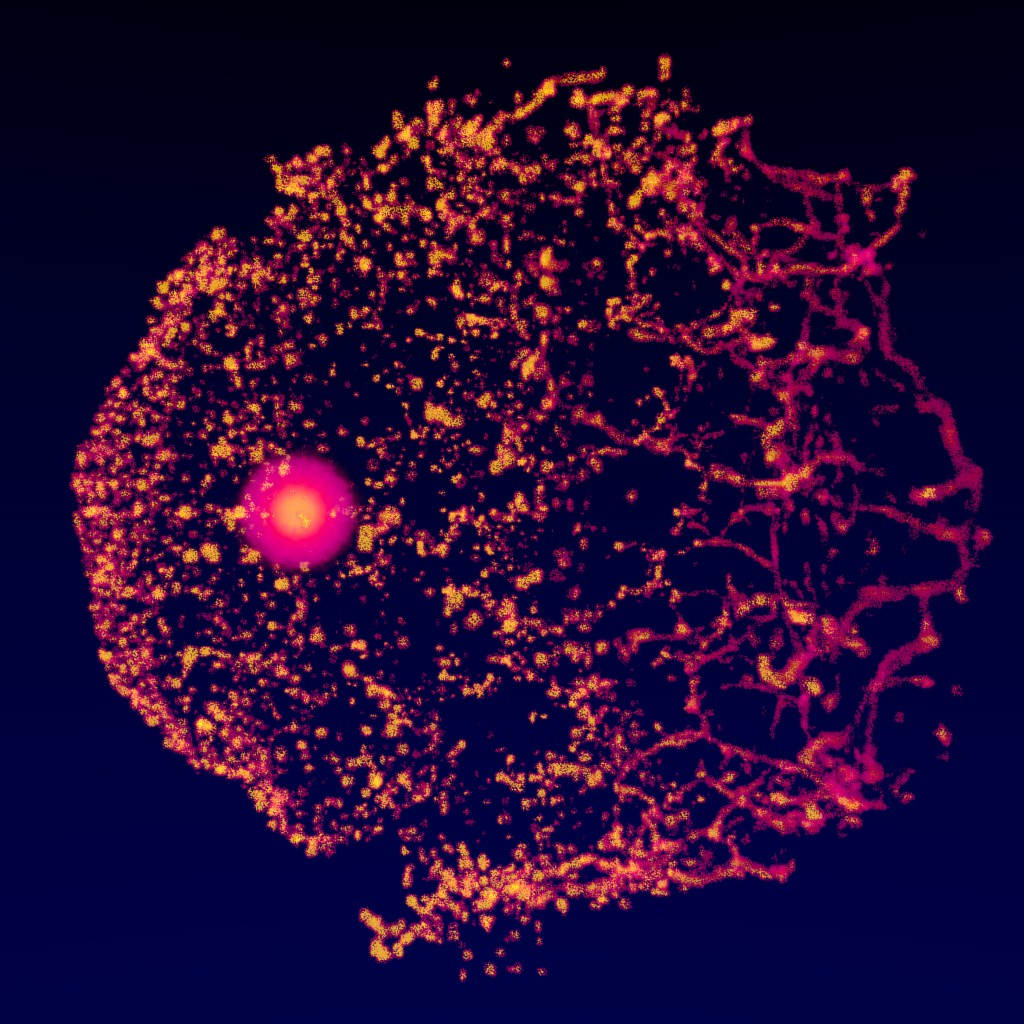}
 \caption{Stellar winds from a Wolf-Rayet star moving through the galactic center \citep{cuadra08}. Simulation data provided by J. Cuadra. The output was converted from an SPH simulation as described in section \ref{sec_convert_particle}. The image was post-processed to brighten the colors, and also the Blend Sky option of the World properties was used to add the background colors.}
 \label{ExampleB}
\end{center}
\end{figure}

The Voxel Data visualization described in this work was tested for Gadget2 and FARGO3D simulation outputs. The Gadget2 code uses the lagrangian formalism with the Smoothed Particle Hydrodynamics (SPH) method, while FARGO3D uses the eulerian formalism and support cartesian, cylindrical, and spherical grids.\\ \indent 
The image shown in Figure \ref{ExampleA} was rendered using FARGO3D outputs provided by S. Perez. The figure shows a protoplanetary disk with a giant planet embedded, and the between the planet and the gas. The planet gravity opens a gap along its orbit and excites the spiral wakes that propagate through the disk. Central and surrounding stars were included for aesthetic reasons. 
The available video animation\footnote{\href{https://www.youtube.com/watch?v=ahf3J_6iv3s}{www.youtube.com/watch?v=ahf3J\_6iv3s}} shows the gap opening process and the circumplanetary disk that surrounds the planet.\\ \indent 
The image shown in Figure \ref{ExampleB} was rendered using Gadget2 outputs provided by J. Cuadra. The figure shows the stellar winds emitted by a Wolf-Rayet star as it moves through the galactic center, and how the ejected material form clumps by cooling in the interstellar medium.
The available turn-around animation\footnote{\href{https://www.youtube.com/watch?v=IQh-rvOLtdQ}{www.youtube.com/watch?v=IQh-rvOLtdQ}} also shows that the clumps are distributed in a \textquoteleft parabolic shell\textquoteright, rather than through all the space (as could be misinterpreted by using  a projection view).\\ \indent
In both examples (images and videos) we illustrated different animation techniques that help to highlight the particular features of each simulation, such as the circumplanetary region or the clump distribution. 
We also encourage the reader to explore more features of the software, in order to find the most effective way to present its own work, and guide its audience through the data visualization.

\section{Summary}
We have discussed how to use the Volumetric Rendering features of the software Blender to display the results of astrophysical simulations. 
We have described the Voxel Data formats, which are binary files (or image sequences) used to import a datacube into the Blender interface, and discussed the advantages and disadvantages of each one in terms of flexibility and memory usage. 
In this context, we have proposed algorithms to convert simulation outputs (either grid or particle based) into a voxel datacube, and provided examples using stellar winds Gadget2 simulations, and protoplanetary disks FARGO3D simulations.\\ \indent 
In the Blender interface we showed the process of setting up a scene, loading the Voxel Data file, and adjusting the colorscale to render an image. We have also presented some basic tools, such as color balance corrections and camera-fly around animation, that may help to create appealing outreach material.\\ \indent 
The whole process is available in our mentioned Youtube tutorial series ``Blender \& Astronomy Tutorial. Using Voxel Data for 
3D visualization''. We want to mention that though this article was focused on representing numerical simulations, the overall procedure is the same if the information can be stored in a datacube.\\ \indent 
The use of Volume Rendering and 3D visualization offers a whole new range of possibilities to perform complementary data analysis and to display our results for the public. Softwares like Blender present many useful tools to accomplish this task. Furthermore, along with the evolution of computer graphics more alternatives will be available, allowing us to create more detailed and effective visualizations.

\section{Acknowledgements}
We would like to thank Jorge Cuadra and Sebastián Perez for providing their simulation data to test this visualization procedure, and Pablo Benítez-Llambay for his collaboration in the development of the FARGO3D to Voxel Data converter. We thank S. Perez, P. Benítez-Llambay, and D. Calderón for their comments in the early version of this paper, and also to the anonymous referee for the comments and suggestions to improve this article.
The author acknowledges financial support from the Millennium Nucleus RC130007 (Chilean Ministry of Economy) and the associated PME project: “MAD Community Cluster”, from FONDECYT grant 1141175, and Basal (PFB0609) grant.
The Geryon clusters housed at the Centro de Astro-Ingenieria UC were used for the SPH calculations of this paper. 
The BASAL PFB-06 CATA, Anillo ACT-86, FONDEQUIP AIC-57, and QUIMAL 130008 provided funding for several improvements to the Geryon clusters. 
The FARGO3D simulations used in this work were performed in the Belka cluster, financed by FONDEQUIP project EQM140101 and housed at MAD/Cerro Calan.

\newpage
\noindent This is an author-created, un-copyedited version of an article accepted for publication in Publications of the Astronomical Society of the Pacific. IOP Publishing Ltd is not responsible for any errors or omissions in this version of the manuscript or any version derived from it.

\end{document}